\documentstyle[amssymb,aps,12pt]{revtex}

\begin{document}
\draft
\author{Sergio De Filippo\cite{byline}}
\address{Dipartimento di Fisica ''E. R. Caianiello'', Universit\`{a} di Salerno\\
Via Allende I-84081 Baronissi (SA) ITALY\\
Tel: +39 089 965 229, Fax: +39 089 965 275, e-mail: defilippo@sa.infn.it\\
and \\
Unit\`{a} INFM Salerno}
\date{\today}
\title{Emergence of classicality in non relativistic quantum mechanics through
gravitational self-interaction}
\maketitle

\begin{abstract}
An explicit dynamical model for non relativistic quantum mechanics with an
effective gravitational interaction is proposed, which, as being well
defined, allows in principle for the evaluation of every physical quantity.
Its non unitary dynamics results from a unitary one in a space with twice as
many degrees of freedom as the ordinary ones. It exhibits a threshold for
the emergence of classical behavior for bodies of ordinary density on rather
long times or instantaneously, respectively at around $10^{11}$ or $10^{20}$
proton masses.
\end{abstract}

\pacs{03.65.-w, 03.65.Bz}

\bigskip

\bigskip

Reconciling gravitation and microphysics\cite{dirac1} has been enduring as
one of the fundamental open theoretical problems since the birth of Quantum
Mechanics (QM), even more challenging after the renormalization of
electrodynamics by Feynman, Tomonaga, and Schwinger\cite{schwinger}. Even
the preliminary questions, in the author's opinion, stay essentially
unanswered as Feynman left them in his lectures on gravitation: ''...maybe
nature is trying to tell us something here, maybe we should not try to
quantize gravity''\cite{feynman}. On the other hand, according to several
authors\cite{bell,ghirardi,pearle1,percival}, the conventional
interpretation of QM is not completely satisfactory due to the dualistic
description it gives for measurement processes and for time evolution of
isolated microsystems.

A possible link between these two issues has been suggested on several
grounds\cite{karolyhazy,diosi0,diosi,penrose,kumar}. The most compelling
reason seems to be the most elementary one. An interaction responsible for a
non unitary dynamics, leading to wave function collapse and then implying
localization, would inevitably be able to inject energy and consequently, in
principle, to induce internal excitations in atomic systems. The excitation
rate is minimized and presumably made compatible with experimental data \cite
{squires,ring} if the main coupling is with the center of mass motion, which
is the case for the gravitational interaction only.

On the other hand the most relevant implications of the measurement problem
involve low energy physics, which suggests that a solution should first be
looked for within the usual setting of the latter. One of its cornerstones
consists in starting from non relativistic particles with instantaneous
action at a distance, which in our case means Newton interaction. Since our
proposal aims at addressing even the issue of the localization of a single
isolated particle, it has to include self-interaction. If this is done
within the usual setting then, on one side it implies some nonlinear
generalization of quantum mechanics, and, on the other side it does not
introduce further degrees of freedom, which only can lead to an effective
non unitary evolution of the original ones.

We intend to show that the two unwanted features mentioned above can be
avoided simultaneously, staying within the framework of finitely many
degrees of freedom and simply duplicating them. Each particle is replaced by
(at least) two metaparticles, say a green one and a red one; in the absence
of gravitational interactions the green and the red metaworlds are
dynamically decoupled and each one is a replica of one and the same
gravitationless world. A Newton interaction is introduced between
metaparticles of different color only. The physical (meta)state space, which
is dynamically invariant, is restricted to metastates where metaparticles
are in couples of different colors, and within each couple the two
metaparticles share the same wave function. Ordinary particles can be
identified say with green metaparticles and ordinary states with the ones
obtained by tracing out red metaparticles. Accordingly the Newton
interaction between metaparticles belonging to different couples gives rise
to an effective gravitational attraction between metaparticles of the same
metaworld, whereas the one between a green metaparticle and its red partner
may be involved in wave function localization only. Such a model represents
the most economical way gravitation can induce a non unitary evolution,
namely representing it as an interaction between physical and unobservable
degrees of freedom.

To be specific, let $H[\psi ^{\dagger },\psi ]$ denote the second quantized
non-relativistic Hamiltonian of a finite number of particle species, like
electrons, nuclei, ions, atoms and/or molecules, according to the energy
scale. For notational simplicity $\psi ^{\dagger },\psi $ denote the whole
set $\psi _{j}^{\dagger }(x),\psi _{j}(x)$ of creation-annihilation
operators, i.e. one couple per particle species and spin component. This
Hamiltonian includes the usual electromagnetic interactions accounted for in
atomic and molecular physics. To incorporate gravitational interactions
including self-interactions, we introduce complementary
creation-annihilation operators $\chi _{j}^{\dagger }(x),\chi _{j}(x)$ and
the overall Hamiltonian 
\begin{equation}
H_{G}=H[\psi ^{\dagger },\psi ]+H[\chi ^{\dagger },\chi
]-G\sum_{j,k}m_{j}m_{k}\int dxdy\frac{\psi _{j}^{\dagger }(x)\psi
_{j}(x)\chi _{k}^{\dagger }(y)\chi _{k}(y)}{|x-y|},
\end{equation}
acting on the tensor product $F_{\psi }\otimes F_{\chi }$ of the Fock spaces
of the $\psi $ and $\chi $ operators, where $m_{i}$ denotes the mass of the $%
i$-th particle species and $G$ is the gravitational constant. While the $%
\chi $ operators are taken to obey the same statistics as the original
operators $\psi $, we take advantage of the arbitrariness pertaining to
distinct operators and, for simplicity, we choose them commuting with one
another: $[\psi ,\chi ]$ $_{-}=[\psi ,\chi ^{\dagger }]_{-}=0$.

The metaparticle state space $S$ is identified with the subspace of $F_{\psi
}\otimes F_{\chi }$ including the metastates obtained from the vacuum $%
\left| 0\right\rangle =\left| 0\right\rangle _{\psi }\otimes \left|
0\right\rangle _{\chi }$ by applying operators built in terms of the
products $\psi _{j}^{\dagger }(x)\chi _{j}^{\dagger }(y)$ and symmetrical
with respect to the interchange $\psi ^{\dagger }\leftrightarrow \chi
^{\dagger }$, which, as a consequence, have the same number of $\psi $
(green) and $\chi $ (red) metaparticles of each species. In particular for
instance the metastates containing one green and one red $j$-metaparticle
are built by linear combinations of the symmetrized bilocal operators 
\begin{equation}
\Phi _{j}^{\dagger }(x,y)\doteq \psi _{j}^{\dagger }(x)\chi _{j}^{\dagger
}(y)+\psi _{j}^{\dagger }(y)\chi _{j}^{\dagger }(x),
\end{equation}
by which the most general metastate corresponding to one particle states is
represented by 
\begin{equation}
\left| \left| f\right\rangle \right\rangle =\int dx\int dyf(x,y)\psi
_{j}^{\dagger }(x)\chi _{j}^{\dagger }(y)\left| 0\right\rangle
,\;\;f(x,y)=f(y,x).
\end{equation}
This is a consistent definition since the overall Hamiltonian is such that
the corresponding time evolution is a group of (unitary) endomorphisms of $S$%
. If we prepare a pure $n$-particle state, represented in the original
setting - excluding gravitational interactions - by 
\begin{equation}
\left| g\right\rangle \doteq \int d^{n}xg(x_{1},x_{2},...,x_{n})\psi
_{j_{1}}^{\dagger }(x_{1})\psi _{j_{2}}^{\dagger }(x_{2})...\psi
_{j_{n}}^{\dagger }(x_{n})\left| 0\right\rangle ,
\end{equation}
its representation in $S$ is given by the metastate 
\begin{equation}
\left| \left| g\otimes g\right\rangle \right\rangle =\int
d^{n}xd^{n}yg(x_{1},...,x_{n})g(y_{1},...,y_{n})\psi _{j_{1}}^{\dagger
}(x_{1})...\psi _{j_{n}}^{\dagger }(x_{n})\chi _{j_{1}}^{\dagger
}(y_{1})...\chi _{j_{n}}^{\dagger }(y_{n})\left| 0\right\rangle .
\label{initial}
\end{equation}
As for the physical algebra, it is identified with the operator algebra of
say the green metaworld. In view of this, expectation values can be
evaluated by preliminarily tracing out the $\chi $ operators and then taking
the average in accordance with the traditional setting.

While we are talking trivialities as to an initial metastate like in Eq. (%
\ref{initial}), that is not the case in the course of time, since the
overall Hamiltonian produces entanglement between the two metaworlds,
leading, once $\chi $ operators are traced out, to mixed states of the
physical algebra. The ensuing non-unitary evolution induces both an
effective interaction mimicking gravitation, and wave function localization.

In fact, if we evaluate the time derivative of the canonical linear
momentum, for notational simplicity for particles of one and the same type,
we get in the Heisenberg picture 
\begin{eqnarray}
\frac{d\overrightarrow{p}}{dt} &=&-i\hslash \frac{d}{dt}\int dx\psi
^{\dagger }(x)\nabla \psi (x)\equiv \vec{F}+\vec{F}_{G}=  \nonumber \\
&&-\frac{i}{\hslash }\left[ \overrightarrow{p},H[\psi ^{\dagger },\psi ]%
\right] +Gm^{2}\int dx\psi ^{\dagger }(x)\psi (x)\nabla _{x}\int dy\frac{%
\chi ^{\dagger }(y)\chi (y)}{\left| x-y\right| }.  \label{dpdt}
\end{eqnarray}
If $\overrightarrow{p}$ denotes the total linear momentum, i.e. the $x$
integration extends to the whole space, and the expectation value in an
arbitrary metastate vector of $S$ is considered, the gravitational force
vanishes, as it should be for self-gravitating matter, due to the
antisymmetry of the kernel $\nabla _{x}(1/\left| x-y\right| )$ and the
symmetry of the metastates in the exchange $\psi ^{\dagger }\leftrightarrow
\chi ^{\dagger }$. On the other hand, if we evaluate the time derivative of
the linear momentum of a body contained in the space region $\Omega $, the
expectation of the corresponding gravitational force is 
\begin{eqnarray}
\left\langle \left( \vec{F}_{G}\right) _{\Omega }\right\rangle  &\simeq
&Gm^{2}\left\langle \int\limits_{\Omega }dx\psi ^{\dagger }(x)\psi (x)\nabla
_{x}\int\limits_{\Omega }dy\frac{\chi ^{\dagger }(y)\chi (y)}{\left|
x-y\right| }\right\rangle   \nonumber \\
&&+Gm^{2}\int\limits_{\Omega }dx\left\langle \psi ^{\dagger }(x)\psi
(x)\right\rangle \nabla _{x}\int\limits_{R^{3}\backslash \Omega }dy\frac{%
\left\langle \chi ^{\dagger }(y)\chi (y)\right\rangle }{\left| x-y\right| } 
\nonumber \\
&=&Gm^{2}\int\limits_{\Omega }dx\left\langle \psi ^{\dagger }(x)\psi
(x)\right\rangle \nabla _{x}\int\limits_{R^{3}\backslash \Omega }dy\frac{%
\left\langle \psi ^{\dagger }(x)\psi (x)\right\rangle }{\left| x-y\right| }.
\end{eqnarray}
The term referring to body self-interaction above vanishes once again due to
symmmetry reasons, while in the following term long range correlations where
considered irrelevant as usual, and in the final result metastate symmetry
was used once more. As for the center of mass coordinate, of course the
expression of its time derivative does not depend on gravitational
interactions at all.

This shows that the present model reproduces the classical aspects of the
naive theory without red metaparticles and with direct Coulomb-like
interactions between distinct particles only. On the other hand they
disagree as for the time dependence of phase coherences.

Consider in fact in the traditional gravitationless setting a physical body
in a given quantum state whose wave function $\Psi _{CM}(X)\Psi
_{INT}(x_{i}-x_{j})$ is the product of the wave function of the center of
mass and an internal stationary wave function dependent on a subset, for
instance, of the electronic and nuclear coordinates. In particular $\Psi
_{CM}$ can be chosen, for simplicity, in such a way that the corresponding
wave function in our model is itself, at least approximately, the product of
four factors: ${\it i}$)a wave function of the center of metamass, namely $%
(X+Y)/2$, where $Y$ is the center of mass of the corresponding red metabody, 
${\it ii}$)a stationary function of $X-Y$ describing the relative motion and 
${\it iii}$) $\Psi _{INT}(x_{i}-x_{j})$ and ${\it iv}$)its red partner $\Psi
_{INT}(y_{i}-y_{j})$, namely: 
\begin{eqnarray}
&&\Psi _{CM}(X)\Psi _{INT}(x_{i}-x_{j})\Psi _{CM}(Y)\Psi _{INT}(y_{i}-y_{j})
\nonumber \\
&=&\tilde{\Psi}_{CM}(\frac{X+Y}{2})\tilde{\Psi}_{INT}(X-Y)\Psi
_{INT}(x_{i}-x_{j})\Psi _{INT}(y_{i}-y_{j}).  \label{localized}
\end{eqnarray}
In Eq. (\ref{localized}) $\Psi _{INT}(x_{i}-x_{j})$ and its red partner are
obviously still stationary to an excellent approximation for not too large a
body mass $M$, since they are determined essentially from electromagnetic
interactions only. As to $\tilde{\Psi}_{INT}(X-Y)$ we choose it as the
ground state of the relative motion of the two interpenetrating metabodies,
which is formally equivalent to the plasma oscillations of two opposite
charge distributions. The corresponding potential energy, if the body is
spherically symmetric and not too far from being a homogeneous distribution
of radius $R,$ can be approximated for small relative displacements, on
purely dimensional grounds, by 
\begin{equation}
U(X-Y)=\frac{1}{2}\alpha \frac{GM^{2}}{R^{3}}\left| X-Y\right| ^{2},
\end{equation}
where $\alpha \sim $ $10^{0}$ is a dimensionless constant. That means that
the relative ground state is represented by 
\begin{equation}
\tilde{\Psi}_{INT}(X-Y)=\left( 2\Lambda ^{2}\pi \right) ^{-3/2}\exp \frac{%
-\left| X-Y\right| ^{2}}{2\Lambda ^{2}},\;\;\Lambda ^{2}=\frac{\hslash }{%
\sqrt{\alpha GM^{3}/R^{3}}}.  \label{gaussian}
\end{equation}
Then, if we choose $\Psi _{CM}(X)\propto \exp \left[ -(X/\Lambda )^{2}\right]
$, we get 
\begin{equation}
\Psi _{CM}(X)\Psi _{CM}(Y)=\tilde{\Psi}_{INT}(X-Y)\tilde{\Psi}_{INT}(X+Y),
\end{equation}
with $\tilde{\Psi}_{INT}$ as in Eq. (\ref{gaussian}). In particular for
bodies of ordinary density $\sim 10^{24}m_{p}/cm^{3}$, where $m_{p}$ denotes
the proton mass, one gets 
\begin{equation}
\Lambda \sim \sqrt{\frac{m_{p}}{M}}cm,
\end{equation}
which shows that the small displacement approximation is acceptable already
for $M\sim 10^{12}m_{p}$, when $\Lambda \sim 10^{-6}cm$, whereas the body
dimensions are $\sim 10^{-4}cm$.

If - in the traditional setting - we now consider at time $t=0$, omitting
the irrelevant factor $\Psi _{INT}$, a superposition 
\begin{equation}
\frac{1}{\sqrt{2}}\left[ \Psi _{CM}(X)+\Psi _{CM}(X+Z)\right] ,
\end{equation}
the corresponding density matrix, before tracing out the red particle, is
represented in our model by 
\begin{eqnarray}
&&\frac{1}{4}\left[ \bar{\Psi}_{CM}(X^{\prime })+\bar{\Psi}_{CM}(X^{\prime
}+Z)\right] \left[ \bar{\Psi}_{CM}(Y^{\prime })+\bar{\Psi}_{CM}(Y^{\prime
}+Z)\right]  \nonumber \\
&&\left[ \Psi _{CM}(X)+\Psi _{CM}(X+Z)\right] \left[ \Psi _{CM}(Y)+\Psi
_{CM}(Y+Z)\right] .
\end{eqnarray}
For not too long times and $Z\gtrsim 2R$, the main effect of time evolution
is due to the energy difference 
\begin{equation}
E_{BIND}\simeq -GM^{2}/R\simeq 10^{-47}\left( \frac{M}{m_{p}}\right)
^{5/3}erg
\end{equation}
between products $\Psi _{CM}(X)\Psi _{CM}(Y)$, $\Psi _{CM}(X+Z)\Psi
_{CM}(Y+Z)$ corresponding to interpenetrating metabodies and $\Psi
_{CM}(X)\Psi _{CM}(Y+Z)$, $\Psi _{CM}(X+Z)\Psi _{CM}(Y)$, where the
gravitational interaction is irrelevant. After tracing out the red
metaparticles, we get at time $t$ the density matrix 
\begin{eqnarray}
&&\frac{1}{2}\bar{\Psi}_{CM}(X^{\prime })\Psi _{CM}(X)+\frac{1}{2}\bar{\Psi}%
_{CM}(X^{\prime }+Z)\Psi _{CM}(X+Z)  \nonumber \\
&&+\frac{1}{2}\left[ \bar{\Psi}_{CM}(X^{\prime })\Psi _{CM}(X+Z)+\bar{\Psi}%
_{CM}(X^{\prime }+Z)\Psi _{CM}(X)\right] \cos \frac{E_{BIND}t}{\hslash },
\end{eqnarray}
leading to the emergence of classical behavior as soon as coherences for a
macroscopic body are unobservable due to their time oscillation. If for
instance we consider a body of $10^{21}$ proton masses, we get a frequency $%
E_{BIND}/\hslash \sim 10^{15}\sec ^{-1}$, corresponding to a length $\hslash
c/E_{BIND}\sim 10^{-5}cm$, much less than the radius $R\sim 10^{-1}cm$. This
shows that in such a case these coherences are totally unobservable as, in
order to detect them, a measurement time far lower than the time needed for
the light to cross the body would be needed.

Of course when the particle mass is not large enough for the oscillations to
hide coherences, they may decrease in time due to the difference in
spreading between the cases of interpenetrating and separated metabodies. In
the former case the spreading affects only the wave function of $X+Y$, while
in the latter the two gaussian wave packets for the two metabodies have
independent spreading. Since in general a gaussian wave packet $\propto \exp
[-x^{2}/(2\lambda ^{2})]$ spreads in time into a wave packet $\propto \exp
[-x^{2}/(2\lambda ^{2}+2i\hslash t/m)]$, where $m$ is the particle mass, the
typical time $\tau $ for the spreading to affect coherences is quite long: 
\begin{equation}
\tau \sim \frac{M\Lambda ^{2}}{2\pi \hslash }\sim 10^{2}\sec .
\end{equation}

It is worthwhile remarking that, while the expression of the localization
length $\Lambda $ in Eq.\ref{gaussian} holds only for bodies whose mass is
not lower than say $10^{12}$ proton masses, another simple case corresponds
to masses lower than $10^{10}$ proton masses, where the relative motion
between the two metabodies in their ground state has not the character of
plasma oscillations any more, and they can be considered approximately as
point metaparticles. Their ground state wave function then becomes the
hydrogen-like wave function 
\begin{equation}
\Psi (X-Y)\propto e^{-\left| X-Y\right| /a},\;\;a=\frac{2\hslash ^{2}}{GM^{3}%
}\sim 10^{25}\left( \frac{M}{m_{p}}\right) ^{-3}cm,
\end{equation}
which shows that the gravitational self-interaction between the green
metabody and its red partner can be ignored for all practical purposes at
the molecular level.

As a result of the previous analysis, according to the present model,
fundamental decoherence due to gravitation is not expected to hide the
wavelike properties of particles even much larger than fullerene\cite
{fullerene,fullerene1}, while it could still play a crucial role with
reference to the measurement problem in QM. While in fact environment
induced decoherence\cite{zurek} can make it very hard to detect the usually
much weaker effects of fundamental decoherence, it cannot go farther than
produce entanglement with the environment. If and why such entangled states
should collapse is outside its scope.

As to the relationship between our model and Einstein equations of the
gravitational field, the viewpoint we adhered to here is that the latter may
be presumably only a large scale manifestation of a fundamental theory that
may well be out of reach, and whose possible non relativistic limit is the
object of the present letter.

In conclusion we would like to stress that a threshold for
gravitationally-induced wave-function localization was already hinted
elsewhere\cite{diosi0,diosi,kumar,SDF} on heuristic, and primarily
numerological, grounds. However this is the first time, in the author's
knowledge, that a detailed dynamical model is proposed, which, while
reproducing the classical aspects of the gravitational interaction, does it
as a result of a non unitary evolution and may then account for the
emergence of classical behavior. While much remains to be done in order to
analyze more general states and their evolution under the concurrent action
of electromagnetic and gravitational interactions, the model allows in
principle to answer all physically meaningful questions by explicit
evaluation. In particular such a model for fundamental decoherence makes it
possible to look for physical subalgebras having (fundamentally)
decoherence-free states\cite{SDF1}.\newpage Acknowledgments - Financial
support from M.U.R.S.T., Italy and I.N.F.M., Salerno is acknowledged


\begin{references}
\bibitem[*]{byline}  Phone: +39 89 965229; FAX: +39 89 965275; electronic
address: defilippo@sa.infn.it

\bibitem{dirac1}  P.M.A. Dirac, {\it Lectures on Quantum Mechanics}. Belfer
Graduate School of Science. Yeshiva University, NY (1964).

\bibitem{schwinger}  {\it Selected Papers on Quantum Electrodynamics}, J.
Schwinger ed., Dover Pubs. Inc., NY (1958).

\bibitem{feynman}  R.P. Feynman, F.B. Moringo and W.G. Wagner, 1962-3
Lectures on Gravitation. California Institute of Technology (1973), p.12.

\bibitem{bell}  J. S. Bell, in {\it Sixty-Two Years of Uncertainty}, A. I.
Miller ed., Plenum, NY (1990), p.17, refs. therein.

\bibitem{ghirardi}  G. C. Ghirardi, A. Rimini and T. Weber, Phys. Rev.D {\bf %
36}, 3287 (1987).

\bibitem{pearle1}  P. Pearle, in {\it Open Systems and Measurement in
Relativistic Quantum Theory}, H.P. Breuer and F. Petruccione Eds.
(Springer-Verlag, Berlin, 1999), refs. therein.

\bibitem{percival}  I.C. Percival, Proc. R. Soc. Lond. A {\bf 451}, 503
(1995).

\bibitem{karolyhazy}  F. Karolyhazy, A. Frenkel and B. Lukacs B., in {\it %
Quantum Concepts in Space and Time, }R. Penrose and C.J. Isham Eds.
Clarenden, Oxford Science Publications, Oxford (1986), refs. therein.

\bibitem{diosi0}  L. Diosi, Phys. Lett. A {\bf 105}, 199 (1984).

\bibitem{diosi}  L. Diosi, Phys. Rev. A {\bf 40}, 1165 (1989), refs. therein.

\bibitem{penrose}  R. Penrose, Gen. Rel. Grav. {\bf 28}, 581 (1996).

\bibitem{kumar}  D. Kumar, V. Soni, Phys. Lett. A {\bf 271}, 157 (2000).

\bibitem{squires}  P. Pearle, E. Squires, Phys. Rev. Lett. {\bf 73}\ 1
(1994).

\bibitem{ring}  P. Pearle, J. Ring, J. I. Collar and T. Frank, Found. Phys. 
{\bf 29}, 465 (1999).

\bibitem{fullerene}  A.I.M Rae, Nature, {\bf 401} (6754),\ 651 (1999).

\bibitem{fullerene1}  M. Arndt et al., Nature, {\bf 401} (6754), 680 (1999).

\bibitem{zurek}  J.P. Paz, W.J. Zurek, LANL quant-ph/0010011, refs. therein.

\bibitem{SDF}  S. De Filippo, LANL\ quant-ph/0007078.

\bibitem{SDF1}  S. De Filippo, Phys. Rev. A {\bf 29}, 052307 (2000).
\end{references}
\end{document}